# Control of a 2-DoF robotic arm using a P300-based brain-computer interface

Golnoosh Garakani[1], Hamed Ghane[2] and Mohammad Bagher Menhaj[3]

[1] Department of Electrical engineering, Tehran University, Tehran, Iran
[2] Department of Electrical engineering, Islamic Azad University, Bandar Anzali, Iran
[3] Department of Electrical engineering, Amirkabir University of technology, Tehran, Iran

**Abstract:**

In this study, a novel control algorithm, based on a P300-based brain-computer interface (BCI) is fully developed to control a 2-DoF robotic arm. Eight subjects including 5 men and 3 women perform a 2-dimensional target tracking in a simulated environment. Their EEG (Electroencephalography) signals from visual cortex are recorded and P300 components are extracted and evaluated to perform a real-time BCI-based controller. The volunteer's intention is recognized and will be decoded as an appropriate command to control the cursor. The final processed BCI output is used to control a simulated robotic arm in a 2-dimensional space. The results show that the system allows the robot's end-effector to move between arbitrary positions in a point-to-point session with a desired accuracy. Furthermore, it should be noted that the proposed approach is suitable for BCI control applications. This model is tested and compared on the Dataset II of the BCI competition. The best result is obtained with a multi-classifier solution with a recognition rate of 97 percent, without channel selection before the classification.

## 1. Introduction

"Pay attention" to something is an act that is frequently performed in daily life. This behavior occurs at the spontaneous electrical activity EEG by the appearance of a wave called P300. This wave was first reported in 1965 [1]. It appears as a positive deflection in the EEG signals approximately 250−500 *ms* following the presentation of a rare, deviant or target stimulus. P300 has been widely utilized in studies on brain activity disorders [2], [3], memory illusion and lie detection [4], [5], [6] and as the input of a BCI system [7]. In order to detect the P300 wave in noisy EEG signals, feature extraction methods combined with powerful classifiers are generally used in order to extract hidden information from such signals and classify them accurately. Utilizing electrical activity of neuron cortex ensemble to control a robotic arm was an interesting goal that attracted researchers to study about BCI. Generally speaking, BCI makes a direct connection between man and an external tool [8]; however, for practical usages of this system, BCI can be defined as a system that establishes a connection between a man and his surrounding environment, which can be practical through the brain [9].

Nowadays, the interface of brain and computer is a wide spreading topic; On the other hand, deleting this interface and installing a chip inside the human brain can be potentially harmful and only a few experiments have been done on the mice [10], [11] , [12]. These systems can be used in various applications such as helping patients who are relatively mentally healthy but have moving troubles, like patients with Amyotrophic lateral Sclerosis (ALS) and Spinal Cord Injury (SCI) or those with disorders that impair the movement of their organs in connection with the environment [13], [14]. Also, BCI systems can be used to simplify connection of user with other objects and devices such as gaming consoles and smart phones [15].

Designing a proper BCI experiment that has high speed and accuracy in detecting P300 is very important. Generally, visual tests that are designed to extract P300 from EEG can be categorized in two patterns. The first pattern is called the oddball paradigm. In this paradigm, there is a 6×6 matrix of English alphabet letters and numbers that each row and column of this matrix is switched on and off randomly. The signal is typically measured most strongly by the electrodes covering the parietal lobe. The presence, magnitude, topography and emergence time of this signal are often used as metrics of cognitive function in decision making processes. The detection of a P300 wave is equivalent to the detection of where the user was looking 300 ms before its detection. In a P300 speller, the main goal is to detect the P300 peaks in the EEG, accurately and instantly. The accuracy of this detection will ensure a high information transfer rate between the user and the machine [8], [10].

Second pattern is used in applications such as choosing the desired direction, 2-D cursor control [17], [18], games [19], and general applications in which user faces some choices. In this pattern, There are some bulbs and each one indicates one choice and are randomly switched on and off in each round. The attention of user to one of these bulbs makes P300 detectable in EEG. This pattern is applied in the paper and then the output of the BCI is used to actuate the end effecter of a simulated robotic arm.

Here, the EEG signals are used to control a simulated 2-degreesof-freedom (DoF) robotic arm in a point-to-point real time session. The robot consists of two revolute joints, makes it possible for the end-effector to move across a plane. The output of BCI is classified as four main decisions that make the robot to move in four main directions: up, down, left and right. The output of BCI is fed into a pre-processing unit and this unit reads the last brain decision and translates it into a reference position for robotic arm. Eventually, the end-effector moves to follow user's directional orders.

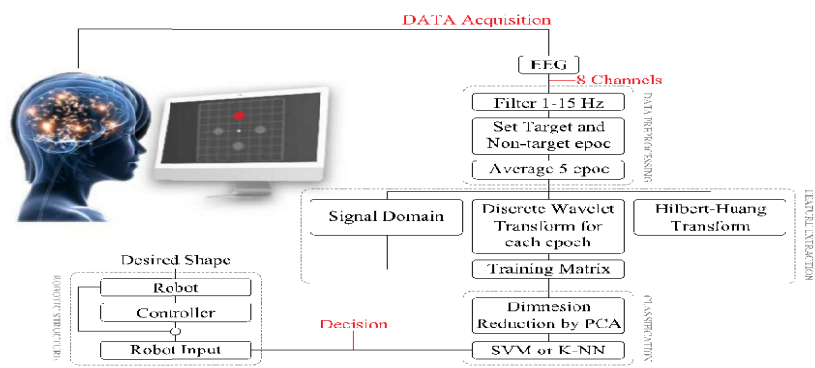

Fig. 1. : **The procedure of controlling a robot by BCI system.**

The whole BCI task is performed in four steps as shown in Fig. 1. First step is to design a test associated with the final goal. The second one is to record the EEG signals followed by a prompt and accurate detection of P300 waves. Third is to process the recorded data. Since brain signals are of low domains, the event related potentials (EPRs) have a small signal to noise ratio (SNR); therefore, by applying a band-pass filter almost all noises could be removed. Then through a Wavelet and Hilbert transforms, features are extracted. In order to detect the subject command, the input data are classified utilizing K-nearest neighbors (KNN) and support vector machine (SVM) classifiers [24], [25], [26]. Finally the forth step is to establish a protocol in order to obtain an efficient connection between the brain and external tools.

The rest of the paper is organized as follows: Section 2 describes recording data process and the primary key factors to improve the results. Section 3 focuses on the preprocessing and filtering the recorded data. Section 4 discusses about EEG signal feature extraction and section 5 presents the classification algorithms. Section 6 proposes the robotic structure and control algorithm as the final application. Simulation results are shown in section 7. Finally, the conclusion is presented in Section 8.

## 2. Data Acquisition

In this section the protocol of designing the task and performance of the recording device are introduced.

### 2-1- Task design

Task design is an important part of each BCI experiment. The more complicated tasks can result more desirable outputs, however executing this complicated task can be very difficult in practice. Therefore, there should be a tradeoff between simplicity of the task and accuracy of the outputs. In addition, factors such as age, gender, education and every other feature of subjects which could potentially affect the results should be carefully considered. Besides, an isolated location, barren of any distraction and noise distortion, should be assigned for the task.

In this study the general models which are in the form of selecting desired direction pattern to record P300, have been evaluated to design the task. Based on the information about time of the task and distance of the bulbs given from the protocols in [27], [28], the primary task has been designed and has performed on subjects. Due to the practical downsides, some correction like tuning the bulbs' distance have been applied after repeating the task several times. The task has carried out on 8 subjects (3 women and 5 men) and the essential data for the research has been obtained. In order to achieve best results, the primary test data has been ignored and the final protocol data has been used.

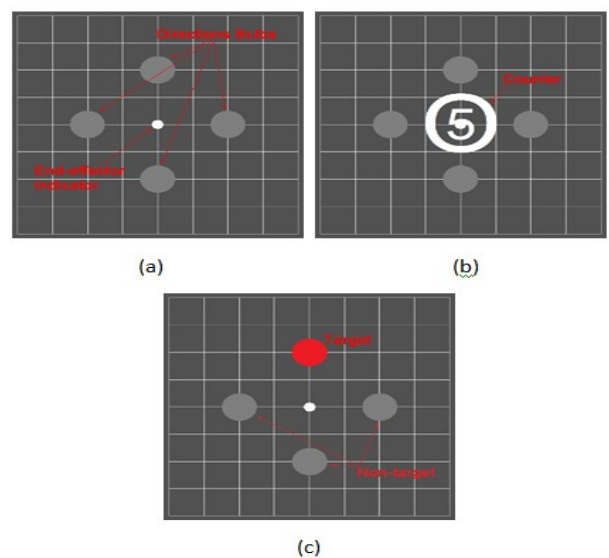

Fig. 2. : **The designed GUI for controlling robotic arm (a) four bulbs as the 4 directions, (b) beginning the task with a counter, (c) target and non-target definition.**

First, four bulbs are shown on *XY* axes with the same distance on a gridded screen as shown in Fig. 2a. The bulbs that are labeled by 1, 2, 3 and 4 represent four main directions. In order to make subjects focus with all their attentions, the experiment began on the screen with 5 seconds countdown that is shown in Fig. 2b. First, a bulb will be randomly switched on and remain on for 100 *ms* while the other three bulbs are off as depicted in Fig. 2c. When the first bulb switches off, the second bulb will be switched on after 150 *ms* and will remain on for the next 100 *ms*. This sequence will be repeated for all bulbs and after 1000 *ms*, the next round will start. To attract the attention of the subjects, they are asked to select one bulb as the target, count and announce the times it switches on. For recognizing selected bulb and consequently selected direction of the subject, this procedure will be repeated 5 epochs. A subject during the data recording is shown in Fig. 3 a. Left laptop displays test while the right one is connected to BCIs to record brain data.

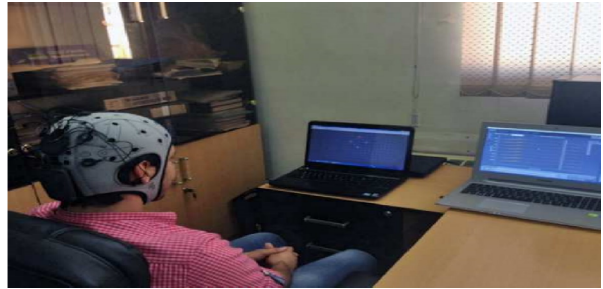

Fig. 3: **EEG signals recording.**

**2-2- Recording Device**

Clinical EEG devices usually have 8, 16, 32, ..., 512 channels and an electrode is implemented to each channel. Typically, electrodes will be placed on the subject's head, which convey the critical potential to preamplifier. Received signals will be amplified and filtered, then they can be recorded for further processing such as frequency spectrum analysis, classification, diagnostic algorithms and analog-to-digital conversion. In this project, the EEG device used to record data is Starstim made in Spain. This device is able to transfer data in eight channels. Transferred data is a $10 \times N$ matrix. First eight columns are the output of channels, the ninth column is the trigger and the last column is the simulation run-time.

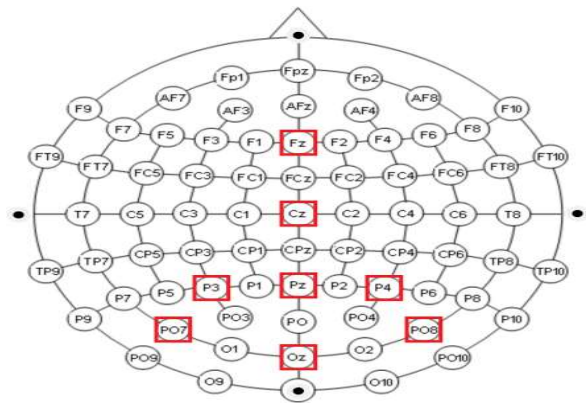

Fig. 4: **The electrodes placement for eight channels.**

The Neuroelectrics Instrument Controller (NIC) is a universal software solution that gives full control of Starstim device. NIC provides a user-friendly interface pack with variety of features. One can manage recordings, launch recording sessions, stream data over the network and receive network triggers with NIC. This software is also able to send the data recorded by Starstim to MATLAB software by using TCP-IP protocol, simultaneously. Brain signals recording require some necessary prerequisites. Primarily, the channels' locations should be specified. This will be done easily by considering the brain's map that is shown in Fig. 4. Then the impedances should be checked in order to record the data with the highest quality.

**2-3- Channel Selection**

Different stimulated brain regions in different subjects, shown in Fig. 5, emphasize the necessity of a crucial channel selection. There are various methods for channel selection depending on the environment and users conditions. Incomplete and improper selection causes difficulties in dimension reduction.

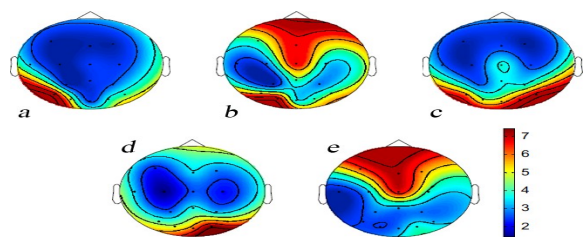

Fig. 5: **Brain activity of five subjects during the test.**

Here, certain bounds of channels in target and non-target signal categories are selected. Then, the correlation for the selected target signals is computed for all channels. Finally, channels with the lowest correlation are ignored. Due to its high speed and accuracy, this method perfectly matches with real-time objectives.

3. **Data Preprocessing**

The next step after data recording is data processing. This step focuses on the quality and statistic assessment of the recorded signals and extraction of their content. In similar studies, the analysis of the training phase has a noticeable effect on the accuracy and efficiency of the results. This emphasizes on the importance of training and testing phase.

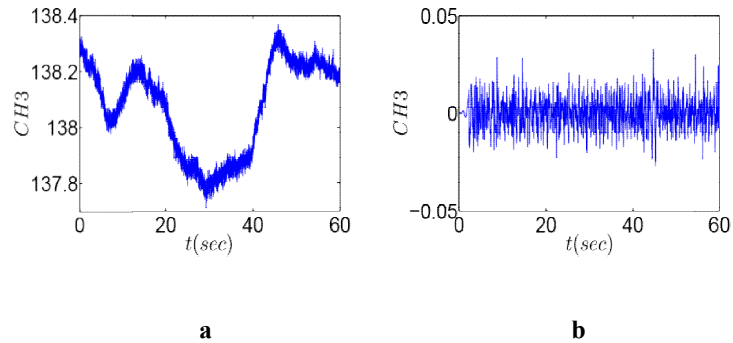

Fig. 6: **Recorded data from channel 3. a: The raw signal, b: Signal after removing trends.**

For removing noise of the signal, it should be filtered by a band-pass filter, but outcome of the filter is zero because of offset trend that is shown in Fig. 6a; therefore, offset trend should be removed in order to prepare signal for further preprocessing. The signal with removed trends is shown in Fig. 6b. Then, the signal is filtered to eliminate all environmental noises like blinking by using the band-pass filter [30]. Sometimes, when the Electrooculography (EOG) and EEG signals amplitude surpass the threshold, there is no need to filter and the extracted parts must be excluded from the process. Considering the fact that P300 components are observable at a frequency band between 1 to 15 HZ, the band-pass filter is designed to maintain these frequencies. After filtering, the signal is extracted to smaller parts based on the beginning of stimulation.

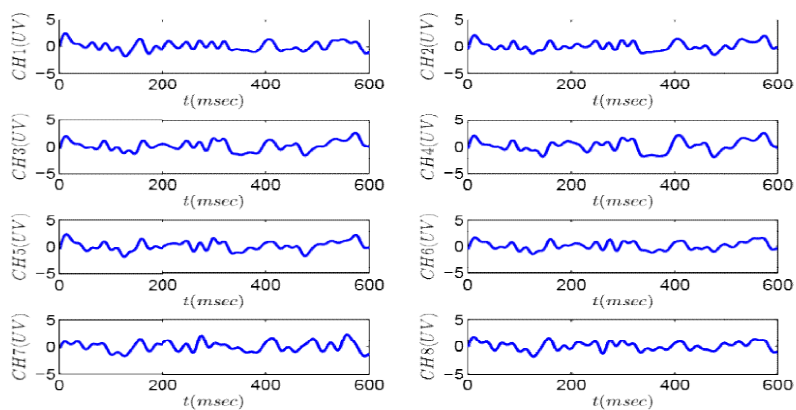

Fig. 7: **The average of five consecutive epochs in range of 600 milliseconds for first subject.**

Increasing the accuracy and simplifying the computing are other advantages of using this frequency range. Afterward, the EEG signal is divided into epochs. Each epoch consists of 600 ms of EEG data after the stimulus onset. In order to increase the SNR and enhance the accuracy of P300 detection, every five successive epochs for each stimulus will be averaged. The output averaged epochs are shown in Fig. 7.

## 4. Feature extraction

A feature usually reflects a distinguishing characteristic, a noticeable measurement or a functional element obtained from data in hand. Extracted features are used to minimize and also simplify the amount of resources needed to represent a massive set of data accurately. Generally, the performance of a BCI system highly depends upon a suitable selection of both features and feature extraction techniques. Here, signal amplitude, Hilbert Huang and wavelet transform coefficients have been assigned as the features. In addition, principal component analysis (PCA) is used to reduce dimension of the feature space.

### 4-1- Hilbert Huang transform

Hilbert Huang transform (HHT) has been essentially developed for the analysis of the nonlinear and non-stationary signals. HHT is a way to decompose a signal into so-called intrinsic (IMF) mode functions along with a trend, and to obtain the instantaneous frequency data [31]. IMF was introduced by Huang et al. [32] as the result of the Empirical Mode Decomposition (EMD). It is a necessary intermediate step toward computing instantaneous frequency through the Hilbert Transform or any other methods. Technically, an IMF is a function that satisfies two conditions: (1) In the whole data set, the number of extrema and the number of zero crossings must be either equal or differ at most by one; and (2) At any point, the mean value of the envelope defined by the local maxima and the envelope defined by the local minima is zero [32]. These features guarantee a well-behaved Hilbert transform.

EMD method is used to decompose the signal into a number of IMF components. Then, by applying the Hilbert transformation to creating analytic signal and obtaining instantaneous frequency and instantaneous amplitude, the initial feature are obtained.

The first IMF has the most similar frequency content and it also has more zero-crossing points comparing to the other functions. This makes the first function, the best alternative to calculate the instantaneous frequency of input signal. In short, the instantaneous frequency calculation using IMF is known as the Hilbert-Huang transform.

**4-2- Wavelet transform**

Among all time-frequency-transformations, wavelets are more effective on ERPs [33]. In ERP analysis, signal phase must be considered and their frequency characteristics require processing in the field of time-frequency. The wavelet coefficients are used in $0-4\ Hz$ and $4-8\ Hz$ frequency bands that are named as Delta and Theta respectively. The quality of decomposition has a direct relation with the compatibility of signal and wavelet transform. In this study, the B-Spline wavelets are used because these types of wavelets are sub-optimal time-frequency localizer. They are also semi-orthogonal and have compact supports. These abilities make them quite popular.

**4-3- Feature space reduction**

Practically, only a small part of features is really important and discriminative. Indeed, the feature space defined on the original signals may contain redundant information that does not have a significant influence on the features categorizing. To preserve the useful information, dimension reduction is applied. This method decreases the computations and also increases the system generalization capacity.

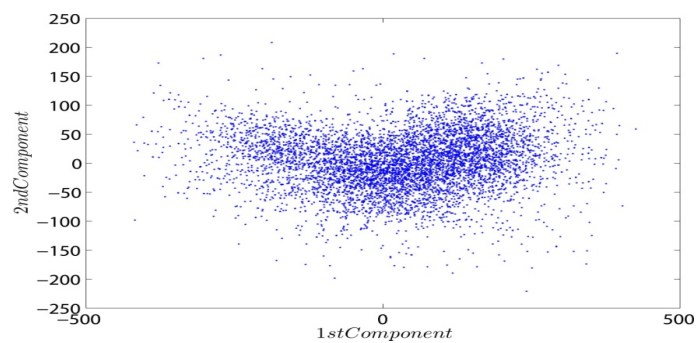

Fig. 8: **Dimension reduction procedure.**

As a scale-invariant method [32], in this paper, PCA is applied for dimension reduction procedure. First, PCA sorts the eigenvalues of training covariance matrix from largest to smallest, then based on the problem definition, only information related to some of the largest eigenvalues are maintained. Finally, using their corresponding eigenvectors, dimension of the secondary features space is reduced comparing to the primary one. Therefore, most of the initial features information will be kept in the secondary space. In Fig. 8, scatter plots for two components with the highest value in PCA are shown. This plots show how much one component is affected by another one.

5. **Classification**

In BCI systems, classifiers are used to organize data based on their extracted features. There are so many methods which have been used for EEG signals classification. The k-nearest neighbors algorithm ( k-NN) and support vector machine (SVM) are two classifiers which are used in this project.

**5-1- Support vector machine ( SVM )**

In P300 BCI research, SVM is regarded as one of the most accurate classifiers [32]. Basic definition of SVM is provided in [35] .The main idea of a linear SVM is to find the separating hyper plane, between two classes such that the distances between the hyper plane and the closest points from both classes are maximal. In other words, we need to maximize the margin between the two classes. Although SVM is generally a two-class classifier, it can be developed for a multi-class condition. Besides the linear SVMs, it can also perform a non-linear classification by using kernel trick. Non-linear SVMs are able to map their inputs into high-dimensional feature spaces. The two dimensional feature space of training data set for first subject is shown in Fig. 9. The target and non-target data are not quite separable. Depending on the subject and the test, there are two or more target groups. As shown in Fig. 9, the targets are in two separate domains of the feature space. These two groups of targets have different characteristics but have to be classified as one target group. This classification is indeed regardless of data distribution; therefore, realizing the training data distribution in the feature space is a necessary task.

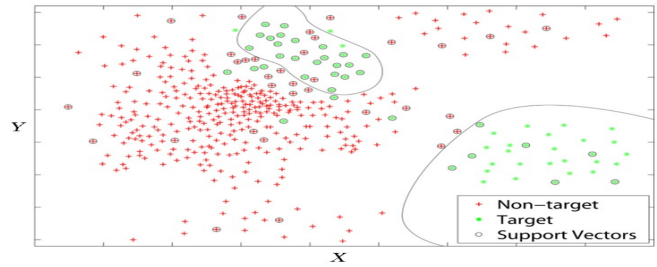

Fig. 9: **Classification of learning data by SVM with optimized Gaussian center for first subject.**

Regardless of using some two-class or a multi-class SVM, a desirable performance is not always guaranteed. So, the k-NN algorithm, as an alternative and possibly better solution is also tried.

**5-2- The k-Nearest Neighbors Algorithm ( k-NN )**

k-NN is a training or supervision algorithm. In general, this algorithm is utilized for two main purposes: 1. the distribution estimation of training density function and 2. data classification according to the training template. The k-NN is used to classify the test data according to the training patterns. This algorithm is compatible with all types of data distribution. As shown in fig. 9, our data are classified in two groups. These two groups aren't so distinct in certain areas. Therefore, it is better to use a multi-class SVM. Choosing between the two-class or multi-class character of the relevant samples related to the first category - which is the group in question- depends on the person or the decided exam. Indeed, using the multi-class SVM requires knowing the methods of data distribution of the training features. On the other hand, applying KNN would eliminate this problem and is compatible with all distribution forms of features.

Using one neighbor does not lead to suitable results and may face with errors, especially at the edge of the borders. Also, using five neighbors may cause same errors. The experiment shows that using three neighbors has the best results in this project. The results for different classifiers considering two sets of data are compared in Fig. 10. The BCI competition II database, Berlin 2002, is used in Fig. 10a. Fig. 10b is based on the recorded data of subjects.

As stated before, it is seen that if we utilize the one neighbor approach, we can't achieve good results in the edges of the two groups and then we might face some errors. On the other hand, if we use 5 or more vicinities in the certain areas, usually close to the edges, we might encounter some other difficulties. Our tests indicate that 3 neighbor will yield the best result.

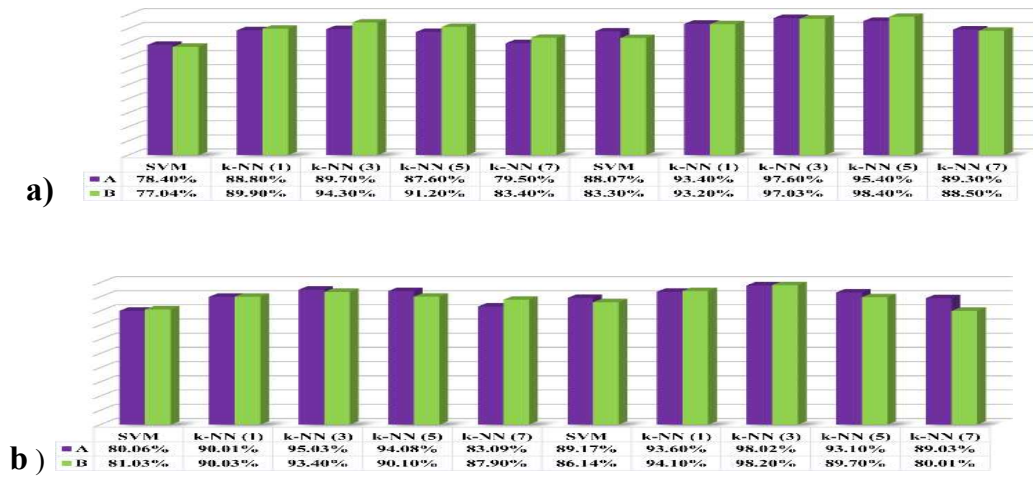

Fig. 10: **Different classifier results for two groups of data sets; (a) BCI competition II, (b) recorded from subjects.**

The comparison results show that the best results come from our database by k-NN classifier.

## 6. Robotic structure

Modeling and control of robotic arm had received tremendous attention in the field of mechatronics over the past few decades and the quest for new development of robot arm control still continues. Here, we are proposing some kind of cognitive robotic control.

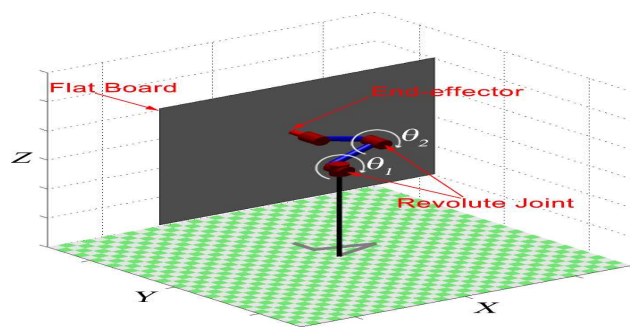

Fig. 11: **A 2-DoF robotic arm.**

In order to evaluate the proposed neuro-cognitive algorithm, the output of BCI is fed to a simulated 2-DoF robotic arm which is shown in Fig. 11. The robot is able to move in a two dimensional space (x-z plane) which provides an ability to do variety of two dimensional tasks including drawing shapes on a flat board, engraving on all kinds of surfaces or grabbing and moving any objects over a flat surface like playing chess or checkers. In this section, dynamics and the control scheme of robotic arm will be presented.

**6-1- Reference input**

To control the robot's end-effector, the output decision from the BCI module is translated as a reference input. Indeed, by recognizing the subject's directional decision (up, down, left or right), robot will follow that movement exactly.

For example, if the initial position of robot's end-effector is $(x,z)$ and the confirmed command from the BCI is detected as the up direction, then the desired position $(x_d, z_d)$ will be $(x, z+h)$ where $h$ is the minimum resolution of the robot's movement. Similarly, for moving down, right and left, the desired position for the robot end-effector will be $(x, z-h)$, $(x+h, z)$ and $(x-h, z)$, respectively. By this method, the subject will be able to control the position of the end-effector at each step in a real-time and point-to-point mode. In the next subsection, the dynamic of the robot is presented.

**6-2- Dynamics of the robotic arm**

The applied simulated robot is a robotic arm with 2 degrees of freedom. Dynamics of the proposed robotic structure is expressed by [36] as

$$M(\underline{\theta})\underline{\ddot{\theta}} + C(\underline{\theta}, \underline{\dot{\theta}})\underline{\dot{\theta}} + \underline{g}(\underline{\theta}) = \underline{u} \qquad (1)$$

where $M(\underline{\theta})$ is the $2 \times 2$ inertia matrix of manipulator. The index $m_{ij}$ describes the effect of acceleration of joint $i$ on joint $j$. The inertia matrix is a symmetric, positive definite matrix, whose indices are configuration dependent. $C$ can be interpreted as the damping term in the model and $\underline{g}$ corresponds to the moment generated by gravitational pull at each link and actuator. $\underline{\theta} = [\theta_1 \quad \theta_2]^T$ is the angular displacement of revolute joints and $\underline{u} = [u_1 \quad u_2]^T$ is the input torque vector to each joint..

**6-3- Inverse dynamics control**

This control algorithm is based on the exact linearization of all nonlinear dynamics of the system. It includes additional terms, actively providing the system with a spring-damper behavior. For tracking a given desired trajectory, (1) can be rewritten as:

$$M(\underline{\theta})\underline{\ddot{\theta}} + \underline{h}(\underline{\theta},\underline{\dot{\theta}}) = \underline{u} \qquad (2)$$

where $\underline{h}(\underline{\theta},\underline{\dot{\theta}}) = C(\underline{\theta},\underline{\dot{\theta}})\underline{\theta} + \underline{g}(\underline{\theta})$. An inverse dynamics linearizing control signal u can be obtained as

$$\underline{u} = M(\underline{\theta})\underline{v} + \underline{h}(\underline{\theta},\underline{\dot{\theta}}) \qquad (3)$$

which will lead to a system of double integrators

$$\underline{\ddot{\theta}} = \underline{v} \qquad (4)$$

The new control signal $\underline{v}$ can now be chosen in such a way that the end effecter position, $\underline{x}$ follows a desired trajectory $\underline{x}_d$. This robotic arm has only two translational degrees of freedom as $\underline{p} = [x \quad y]^T$, i.e. the end effecter orientation is fixed. The end effecter's linear velocity is obtained by the help of Analytical Jacobian as:

$$\underline{\dot{p}} = \begin{bmatrix} \frac{\partial x}{\partial \theta_1} & \frac{\partial x}{\partial \theta_2} \\ \frac{\partial y}{\partial \theta_1} & \frac{\partial y}{\partial \theta_2} \end{bmatrix} = J_A(\underline{\theta})\underline{\dot{\theta}} \qquad (5)$$

here $\underline{p} \in \mathbb{R}^2$ describes the position of the end-effector and $J_A(\underline{\theta})$ is the so called Analytical Jacobian. Indeed, $J_A(\underline{\theta})$ corresponds to the mapping between the joint velocities and linear velocity of end effecter. By time differentiation of (5), the acceleration of end-effector can be expressed by:

$$\underline{\ddot{p}} = J_A(\underline{\theta})\underline{\ddot{\theta}} + \dot{J}_A(\underline{\theta}, \underline{\dot{\theta}})\underline{\dot{\theta}} \qquad (6)$$

where $\dot{J}_A(\underline{\theta}, \underline{\dot{\theta}}) = \frac{\partial^2 \underline{p}}{\partial \underline{\theta}^2}\underline{\dot{\theta}}$. Considering (4) and (6), new control signal $\underline{v}$ can now be designed in such way that $\underline{p}$ will follow a given desired trajectory $\underline{p}_d$ as follows:

$$\underline{v} = J_A^{-1}(\underline{\theta})\left(\underline{\ddot{p}}_d + K_D \underline{\dot{e}} + K_P \underline{e} - \dot{J}_A(\underline{\theta}, \underline{\dot{\theta}})\underline{\dot{\theta}}\right) \qquad (7)$$

with positive definite diagonal matrices $K_P$ and $K_D$ and $\underline{e} = \underline{p}_d - \underline{p}(\underline{\theta})$. The error dynamics can be described by

$$\underline{\ddot{e}} + K_D \underline{\dot{e}} + K_P \underline{e} = 0 \qquad (8)$$

The diagonal positive definite property of matrices $K_P$ and $K_D$ will ensure that the desired trajectory $\underline{p}_d$ will asymptotically be reached. By tuning of the control parameters $K_P$ and $K_D$, different dynamic behavior of the system is obtained. The simulation model of the robot arm with implemented Inverse Dynamics Control is illustrated in Fig. 12.

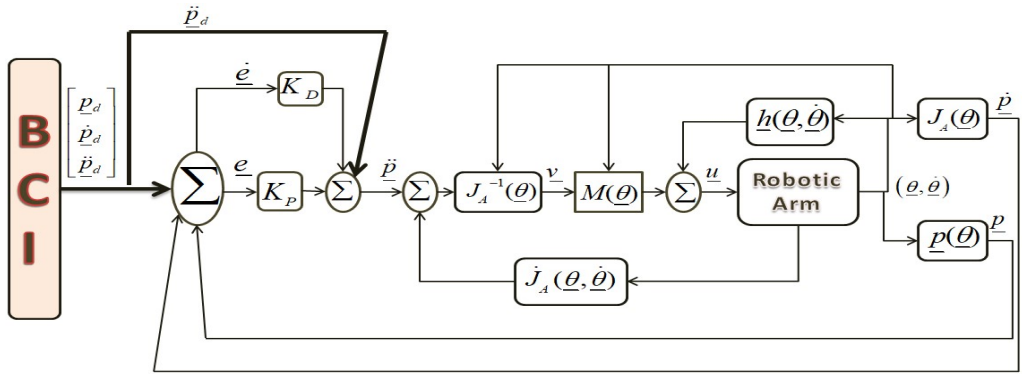

Fig. 12: **Inverse dynamics control.**

## 7. Simulation

In this section, a simulation in MATLAB is given to evaluate the ability of BCI robotic structure. Consider the 2-DoF robotic arm of Fig. 10 which its end-effector is attached to a marker. The task is to draw some desired shapes on a vertical board via the BCI system connected to the user brain. Subject imagines the desired shape in his mind and based on the required movements, at each step, focuses on the corresponding directions bulb in the BCI panel to move the tip of marker (attached to the end-effector). Indeed, BCI reads the confirmed direction, translate it for the robot controller as reference input and finally robot will draw desired shape like a robotic hand.

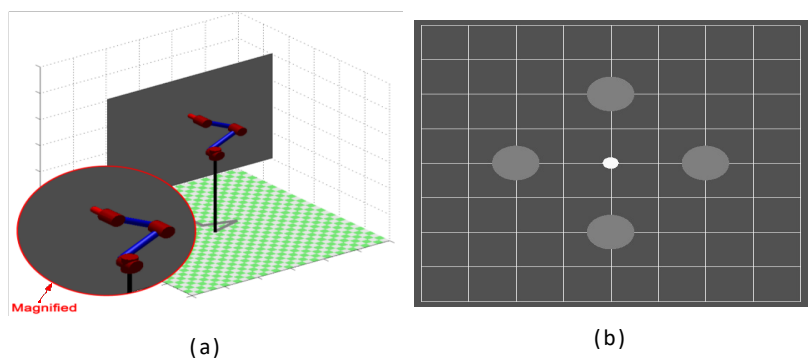

Fig. 13: **Initial state (a) Simulated robotic arm, (b) The BCI panel.**

As an example, drawing a letter like G is considered with more details. The first step is to move the end-effector in up direction. To achieve the first movement, the user has to focus on the upper bulb. By that, the BCI system reads the brains signal, processes it and sends the decision to the robot input unit.

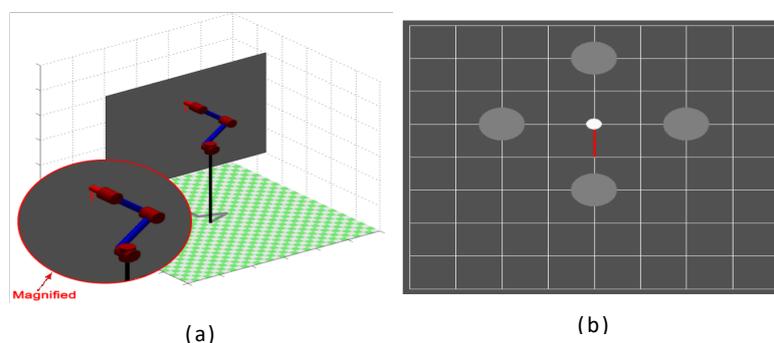

Fig. 14: **After first detected decision ( a ) Simulated robotic arm, (b) The BCI panel.**

Corresponding to the robot fixed resolution, this unit prepares the desired reference input for the robot controller. Then, the end-effector is moved in a way to reach the desired position by the control law. In Fig. 13, the initial position of the robot and BCI panel are shown. After receiving the first detected decision, the robot end-effector follows the controller inputs to meet the desired path as shown in Fig. 14. In the next step of drawing letter G, the subject has to focus on left flashing bulb. Then, by correspondingly focusing on the appropriate flashing bulbs in each step, letter G will be drawn on the board. Considering the counterclockwise hand writing, the sequence of subject focuses is up-left-down-right-up-left. The resulting output of BCI panel is shown in Fig. 15b. In addition, the output trajectory of the simulated robotic arm is depicted in Fig. 15 a.

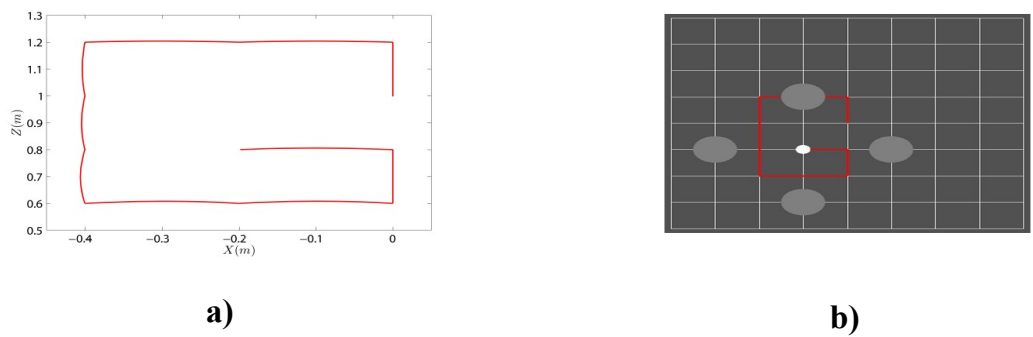

Fig. 15: ( a ) **Final robotic arm trajectory, (b) Final BCI panel display.**

Regarding the data acquisition method, the drawn letter G is sharp cornered and is not smooth. In this study, there was only four directions bulb which finally leads to four distinct directions. Therefore, the robot movements are limited to these four directions which cause hard corners in output shapes.

8. **Conclusion**

In this paper, recording methods, stimulus presentation paradigms, feature extraction and classification algorithms of a P300 based BCI system were studied. We reached high accuracy in classification. The last stage of designing a BCI system is establishing a protocol for fast, accurate and efficient connection between brain, applications and external tools. In order to test the proposed algorithm, BCI system was connected to a 2-DoF robotic arm. The overall structure consisted of BCI as the input and robot end-effector as the output. In this case, user was able to control robot's end-effector position by focusing on desired direction's bulb. According to the mentioned applications of BCI systems, offline use of this type of system is not functional and the main goal is being practical and real time application.

Further work and success of this research would lead to the development of robotic systems that can be used by disabled users, and thus improve their mobility, independence, and quality of life. In this regard, we are currently working on developing some robotic manipulators which are more compatible with BCI systems and improving the quality of attention detection based control schemes.